\documentclass[aps,prd,twocolumn,showpacs,amsmath,preprintnumbers,nofootinbib,superscriptaddress,secnumarabic]{revtex4}
\def\scn#1#2{\section{#1}\lb{#2}}
\def\sscn#1#2{\subsection{#1}\lb{#2}}
 
\usepackage{epsfig,amssymb,amsfonts,verbatim}

\def\bfl{\begin{flushleft}}
\def\efl{\end{flushleft}}
\def\bfr{\begin{flushright}}
\def\efr{\end{flushright}}
\def\bc{\begin{center}}
\def\ec{\end{center}}
\def\be{\begin{equation}}
\def\ee{\end{equation}}
\def\ba{\begin{eqnarray}}
\def\ea{\end{eqnarray}}
\def\baa#1{\begin{array}{#1}}
\def\eaa{\end{array}}
\def\bw{\begin{widetext}}
\def\ew{\end{widetext}}
\def\nn{\nonumber }
\def\lb#1{\label{#1}}
\def\bit{\begin{itemize}}
\def\eit{\end{itemize}}

\def\schrod{Schr\"odinger  }

\def\ntr{n_T}

\def\Sign#1{\, \text{sign}\left(#1\right) }

\newtheorem{Def}{Def.}[section]

\newtheorem{Prp}[Def]{Proposition}

\begin{document}

\preprint{\small J. Phys. B: At. Mol. Opt. Phys. 44 (2011) 195303 [arXiv:1108.0847]}

\title{
Quantum
Bose liquids with
logarithmic nonlinearity:
Self-sustainability
and emergence of spatial extent
%Bose-Einstein Condensate
}

\author{
Alexander V. Avdeenkov}

\affiliation{National Institute for Theoretical Physics (NITheP) and Institute of Theoretical Physics, University of Stellenbosch, Stellenbosch 7600, South Africa}

\affiliation{
Skobeltsyn Institute of Nuclear Physics, Lomonosov Moscow State University,
Leninskie gory, Moscow 119991, Russia}

\author{
Konstantin G. Zloshchastiev}

\affiliation{Department of Physics and Center for Theoretical Physics,
University of Witwatersrand,
Wits 2050, Johannesburg, South Africa}

\affiliation{School of Physics, University of KwaZulu-Natal, Pietermaritzburg Campus,
Private Bag X01 Scottsville, Pietermaritzburg 3209, South Africa}

%\date{~ ~~~~~~~~~~~~~~~~~~~~~~}
%\date{~Received: 12 Jan 2001 [LANL] ~}
%\date{~Received: 26 May 2000 [PRL], 1 June 2000 [LANL] ~}
%\date{Received \today}

%\scriptsize%\footnotesize

\begin{abstract}
The Gross-Pitaevskii (GP)
equation is a long-wavelength approach
widely used to describe the dilute Bose-Einstein condensates (BEC).
However, in many physical situations, such as higher densities,
this approximation unlikely suffices hence one might need models
which would account for long-range correlations and multi-body
interactions.
We show that the Bose liquid described by the logarithmic wave equation
has a number of drastic differences  from the GP one.
It possesses the self-sustainability property: while the free GP condensate
tends to spill all over the available volume the logarithmic one
tends to form a Gaussian-type droplet - even in the absence of an external trapping potential.
The quasi-particle modes of the logarithmic BEC
are shown to acquire a finite size despite the bare particles being assumed point-like, i.e.,
the spatial extent emerges here as a result of quantum many-body correlations.
Finally,
we study the elementary excitations  and demonstrate that
the background density changes the
topological structure of their momentum space
which, in turn, affects
their dispersion relations.
Depending on the density the latter can be of the massive relativistic,
massless relativistic, tachyonic and quaternionic type.
\end{abstract}

\maketitle

%\narrowtext

%\small \newpage

\scn{Introduction}{sec-i}

The Gross-Pitaevskii (GP)
approximation is a long-wavelength theory commonly used
for describing dilute Bose-Einstein condensates (BEC), such as trapped alkali
gases.
As a matter of fact, the GP equation (also known as the cubic \schrod equation)
is the result of few physical
assumptions \cite{Bogoliubov47}: the mean-field approximation, neglected excited states and
multi-body interactions
(three and more particles scatter at one point), two-body interaction being assumed
of the contact type (delta-function),
neglected anomalous contributions to self-energy when using the perturbation
theory \cite{Bel58}.
There is no warranty whatsoever that these assumptions must
hold for all physical examples of BEC.
Indeed, higher densities (common for the Bose liquids made of
noble gases
or nuclear matter under certain conditions \cite{Page:2010aw}),
low-dimensional effects or the presence of mixtures
can
undermine their validity and the theory needs drastic
modifications of a universal form.
Such modifications are usually made in the form of corrections
to the GP equations or, alternatively, to the expression
for the energy functional:
while the energy in the GP approximation is quartic in
$|\Psi |$, $\Psi$ being the wave-function of condensate,
the corrections can be of the sextic and octic power, some examples being given
in Ref. \cite{Schick71}.
Yet, even those corrections may not suffice, e.g., if
higher densities lead to the long-range correlations and multi-body
interactions.
In that case one should account for all the powers of $|\Psi |$ which
implies the usage of the non-polynomial functions which will necessarily
 appear in the wave equations for condensate.
An example of the usage of non-polynomial functions for certain class
of condensates has been
demonstrated, for instance, in Ref. \cite{Salas02}, although
the known examples refer to effectively lower-dimensional models so far.

In this paper we heuristically introduce the condensate
described by the non-linear \schrod equation of a non-polynomial
kind, namely, the logarithmic one (LogSE):
\be\lb{e-xmain}
\left[
\hat{{\cal H}}
-
\beta^{-1} \ln{(a^3 |\Psi|^2)}
\right]
\Psi
= 0,
\ee
where
$\Psi$ refers in general to the complex-valued wave functional,
$\beta$ and $a$
are the parameters of the theory,
and
$\hat{{\cal H}} $ is the operator the form of which
is determined by
a physical setup - for instance, in a non-relativistic theory
$\hat{{\cal H}} =  \hat{\bf H} - i \hbar \partial_t$
where
$
\hat{\bf H} =  \frac{\hat P^2}{2 m} + V_\text{ext} (x)
$
is the Hamiltonian operator.
The physical motivation behind this equation as well as its
unique properties are listed in the Appendix.
Since the pioneering works \cite{BialynickiBirula:1976zp}
this equation received much attention - its
applications were found
not only in the extensions of quantum mechanics,
but also
in quantum optics \cite{buljan03},
nuclear physics \cite{efh85},
transport and diffusion phenomena \cite{transp},
open quantum systems and information
theory \cite{Yasue:1978bx},
effective quantum gravity and physical vacuum
models \cite{Zloshchastiev:2009zw,Zloshchastiev:2009aw}.
Thus, a theory of Bose liquids
%and Bose-Einstein condensates
can
be yet another interesting area of application of LogSE,
certainly worth  being explored.

To begin, let's suppose that we have the BEC of $N$ identical particles
whose wave-function is described by the equation (\ref{e-xmain})
which in our case becomes
\bw
\be\lb{e-becgeneq}
\left[
- i \hbar \, \partial_t
- \frac{\hbar^2}{2 m} \vec \nabla^2
+
V_\text{ext} (\vec x,\, t)
-
\beta^{-1} \ln{(a^3 |\Psi (\vec x, t)|^2)}
\right]
\Psi (\vec x, t)
= 0,
\ee
\ew
where $m$ is the mass of the condensate particle,
$V_\text{ext} (\vec x,\, t)$ is the trapping potential.
Here the notations for $\beta $ are as follows:
our $\beta$
is equivalent to $1/b$ from Ref. \cite{BialynickiBirula:1976zp}
and
has a sign opposite to $\beta$ from the corresponding
part in Ref. \cite{Zloshchastiev:2009aw}.
The particle density of the condensate
is defined as usual:
\be
n = |\Psi (\vec x, t) |^2 =
\left|
\prod\limits_{i=1}^N \phi (\vec x_i, t)
\right|^2
,
\ee
where $\phi (\vec x_i, t)$
describes the single-particle state of the $i$th boson in the fully
condensed state.

For now, we leave aside the question of the microscopical structure
which might lead to Eq. (\ref{e-becgeneq}), be it long-range correlations,
multi-body scattering, fermionic degrees of freedom or anomalous energy contributions,
and study the logarithmic Bose liquid as it is (although, some elements
of the microscopic theory are given below in the
section allotted to the Bogoliubov excitations).
On top of the properties which directly follow from those listed in the Appendix,
this type of condensate has certain unique properties
as a superfluid.

 First of all, let's consider the effective potential energy for the logarithmic BEC.
If one defines the Lagrangian density as
\bw
\be
 L
 %[\Psi^*, \vec\nabla\Psi^*, \partial_t\Psi^*,\vec x_i, t ]
 =
\frac{i\hbar}{2}(\Psi \partial_t\Psi^* - \Psi^*\partial_t\Psi)+
\frac{\hbar^2}{2 m}
|\vec\nabla \Psi|^2
+
V_\text{ext}
|\Psi|^2
-
\beta^{-1}
|\Psi|^2
\left[
\ln{( a^3 |\Psi|^2)} -1
\right]
,
\ee
\ew
then the corresponding Euler-Lagrange equation,
$
\partial_t
\left(
\frac{\partial L}{\partial(\partial_t\Psi^*)}
\right)
+
\vec{\nabla}\cdot
\left(
\frac{\partial L}{\partial (\vec{\nabla} \Psi^*)}
\right)
-
\frac{\partial L}{\partial \Psi^*}=0
,
$
yields Eq. (\ref{e-becgeneq}).
Thus, the effective potential energy density
is given by
\be\lb{e-ftpot}
V_\beta \equiv -
\beta^{-1}
n
\left[
\ln{(n a^3)} -1
\right]
,
\ee
and thus it opens up
(down) and has the local non-zero  minima (maxima) at $n_\text{ext} = a^{-3}$
for the negative (positive) $\beta$.
In what follows we call this potential \textit{logarithmic} -
due to the property
$d V_\beta / d n \propto \ln{(n a^3)}$
which yields the logarithmic term in the \schrod equation.

\begin{figure}[htbt]
\begin{center}\epsfig{figure=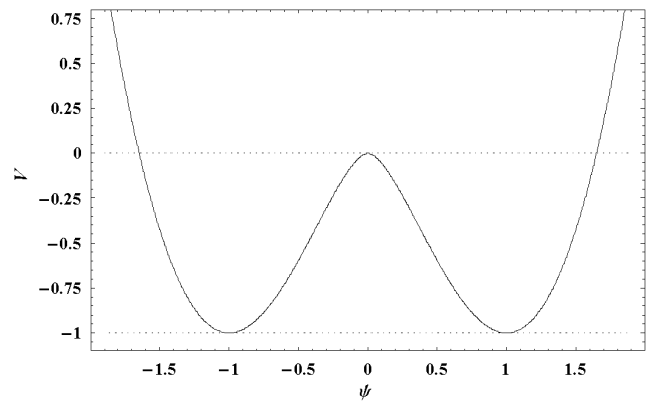,width=  1.05\columnwidth}\end{center}
\caption{
The field-theoretical potential (\ref{e-ftpot}) of the logarithmic BEC 
%(taken with an opposite sign)
in the
dimensionless units: $- \beta a^3$ (vertical axis) and $a^{-3/2}$ (horizontal axis).
}
\label{f:ftpot}
\end{figure}

This potential is non-analytic yet regular at $n = 0$ - while the logarithm itself
diverges there,
the factor $n$ recovers the regularity,
see Fig. \ref{f:ftpot}.
In fact, the potential always has the Mexican-hat shape
if plotted as a function of $\Psi$.
This leads to the drastically new features which will be discussed below.
It is not difficult to check also that the Ginzburg-Landau potential
(the quartic potential supplemented with the chemical-potential quadratic term)
is one of  the perturbative limits of the logarithmic one -
if, for instance, one expands $V_\beta $ in the vicinity of
its local extrema $n_\text{ext}$:
$V_\beta \to - \beta^{-1}
\left[
(a^3 / 2) n^2 - n - 1/(2 a^3)
\right]
+ {\cal O} [(n a^{3} - 1)^3]
$.
Obviously, this expansion only makes sense  when the density
is close to the special value $1/a^{3}$  but
in general the  potential (\ref{e-ftpot}) can not be replaced
by the polynomial one and thus requires a non-perturbative treatment.
%But it has nothing to do with the spontaneous symmetry breaking phenomenon as it keeps such a shape even if one adds  a quadratic term of any sign and value.

For instance,
as long as  the potential (\ref{e-ftpot}) vanishes in the extrema $n_\text{ext}$,
% as it can be  seen from Eq. (\ref{e-becgeneq})
they  describe the equilibrium configurations for ``free'' particles.
These equilibria can be stable or unstable and here appears something
which might look counterintuitive at first sight.
The negative $\beta$ means that the the field-theoretical
potential  opens up and the above-mentioned extrema  are minima   but
the energy functional is \textit{not} bounded from  below \cite{BialynickiBirula:1976zp},
therefore, the stable equilibrium solution can only exist in a  finite volume or in
the additional external potential (``trap'') \cite{Guerrero2010}.
On the contrary, when  $\beta$ is positive then the
potential (\ref{e-ftpot}) opens down but the energy functional is bounded from  below.
An apparent contradiction can be resolved if one recalls that we are dealing with the normalized
wave-function and thus its square (the condensate density $n$) cannot grow arbitrarily large.
The physical consequence is that
the coherence length is always larger than the effective size of the droplet.
In Sec. \ref{sec-becnot} and \ref{sec-bectrap}
we will show that this  leads to one of
the most important features of the Bose liquid
with logarithmic nonlinearity: at positive $\beta$ it can be self-bound and stable (as
to form the droplet-like object)
even without any trapping potential.
%We discuss the details and other features and matter of stability of such a potential .

Second,
the logarithmic term in Eq. (\ref{e-becgeneq})
can be associated with the certain kind of quantum-informational
entropy $S_\Psi \propto - \int |\Psi|^2 \ln{(\Xi |\Psi|^2)} d^3 x$,
measuring the degree of quantum spreading of the BEC as a
collective quantum object, as being described in the Appendix.
The only difference from the interpretation presented there
is that in the BEC case the logarithmic term describes the quantum
information transfer
between the states of the correlated particles which form the many-body state
whereas the temperature $T_\Psi$ is a formal measure
of the collisionless interactions inside the Bose liquid
at zero (thermal) temperature, see the Appendix.

Third,
using the Madelung representation of the wave function
and the hydrodynamic form of the wave equation (\ref{e-becgeneq})
\cite{pn66,Dalfovo:1999zz},
the zero-temperature (collisionless)
equation of state of the
logarithmic BEC in the leading order
with respect to the Planck constant
is described by
a kind of
the Clapeyron-Mendeleev law,
\be\lb{e-eos}
p - p_0 = - (m \beta)^{-1} n + {\cal O} (\hbar^2) \propto T_\Psi n
.
\ee
For comparison, the equation of state for the GP condensate would be
$p \propto n^2$, thus, the logarithmic Bose liquid is more ``ideal'' than the Gross-Pitaevskii one
yet non-trivial.
From here it follows that
the logarithmic BEC is the only one where
the speed of sound (phonons)
%derived from the collisionless equation of state (\ref{e-eos})
does not depend on the density in the leading
order with respect to the Planck constant:
\be\lb{e-ssnd}
c_s =
1/\sqrt{m |\beta |}
+ {\cal O} (\hbar)
.
\ee
%provided the combination $m \beta$ is negative.
%otherwise the real speed of sound cannot be described by this formula.
Meanwhile, the speed of sound for the GP condensate
scales as a square root of density.
Due to this property the logarithmic BEC
can be used to describe or mimic the induced relativity and
several high-energy and gravitational phenomena,
along the lines described in Refs.
\cite{Zloshchastiev:2009aw,Garay:1999sk,Novello:2002qg},
although considering these topics would bring us too
far from the scope of the current paper.

Finally,
in the GP case it is possible
to differentiate between the
``attractive'' and ``repulsive'' interaction depending on a
sign of the coupling constant.
In  our case the logarithm changes
sign
when density goes across the value $a^{-3}$
which results in changing the
sign of the interaction in Eq. (\ref{e-becgeneq})
and the type of the condensate depends not only
on the sign of $\beta $ but also on whether the
value of $n a^3$ is smaller or larger than one.

\scn{The logarithmic BEC in the empty space}{sec-becnot}

While many of the properties discussed in this section
hold also for the condensate in a trap,
it is more clarifying to see them
on the example of the released condensate.
If it were the GP-type Bose gas then
its properties would be nearly trivial:
once the trap is  off the GP condensate
spreads uniformly all over the available space (if it is
repulsive), or shrinks down until the new phase
arises, in the case of attraction.
The behavior of the free logarithmic BEC will be shown to
be much more complex, mostly due to the properties of the logarithm mentioned above.

%\sscn{Ground-state solution}{sec-esgss}

\begin{figure}[htbt]
\begin{center}\epsfig{figure=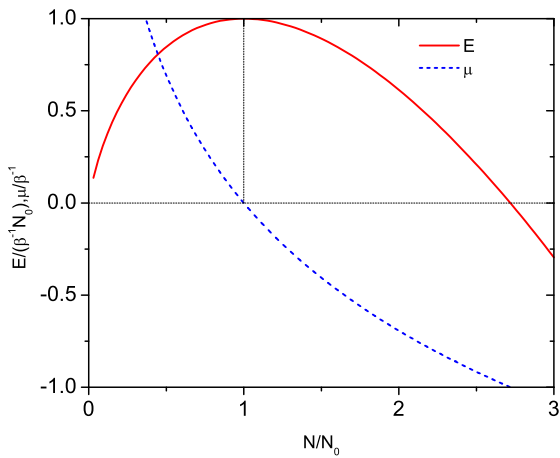,width=  1.05\columnwidth}\end{center}
\caption{
The total energy (solid line) and the chemical potential (dotted line) for the logarithmic BEC with
positive $\beta$
in absence of any trap
as a function of the number of particles.
}
\label{f:e_mu_n}
\end{figure}

If we assume the isotropic case for simplicity
and work in the center-of-mass reference frame
of the condensate then
$\Psi (\vec x, t) = \Psi (|\vec x |) \exp{(- i \mu t / \hbar)}$,
$\mu$ being the chemical potential,
and
the equation (\ref{e-becgeneq}) reduces to
the ordinary differential equation for
the modulus of wave function
$\Psi = \sqrt{n(r)}$
as the function of the radial coordinate only,
\be\lb{e-odenotrap}
\frac{\hbar^2}{2 m}
\Delta \Psi
+
\left[
%\tfrac{1}{2} m \omega^2 r^2
 \mu
+
\beta^{-1} \ln{(a^3 \Psi^2)}
\right]
\Psi
= 0,
\ee
where
$\Delta \Psi = r^{-2} \partial_r (r^2 \partial_r \Psi)$.
Further it will be convenient to introduce
a value of
%length dimensionality,
%$a_\omega = \sqrt{\hbar / m \omega}$,$a_\mu = \hbar / \sqrt{ 2 m \mu }$ and one of
dimensionality of the length squared,
$A_\beta =  \hbar^2 \beta /  2 m $
which
measures the strength of the logarithmic interaction
in terms of length scales,
and
a value of
dimensionality of the energy,
$E_a =  \hbar^2 /  (2 m a^2)$
which
is the de Broglie energy corresponding to the
length scale $a$.
Upon applying the
normalization condition, $\int n d^3 x = 4 \pi \int\limits_0^\infty \Psi^2 r^2 d r = N$,
we deduce that the normalizable solution
%of the gausson type (\ref{e-soltrapnorm})
is only possible
 for positive $\beta$.
The exact expressions for, respectively, the
particle density and the chemical potential are:
\be
n (r)= \Psi^2 =
n_0
e^{-(r/a_\beta)^2}
,
\lb{e-solnotrap}
\ee
\be
\mu =
3 \beta^{-1}
\left[
1 -
\ln{
\left(
\frac{N^{1/3}}{\sqrt{\pi}}
\frac{a}{a_\beta}
\right)
}
\right]
=
-
\beta^{-1}
\ln{
\left(
N / N_0
\right)
}
,
\lb{e-munotrap}
\ee
where
$n_0 = N / (\pi^{3/2} a_\beta^3)
 = (e/a)^3 N/N_0  $,
$a_\beta = \sqrt{A_\beta} = \hbar \sqrt{\beta/ 2 m}$,
and
$
N_0 \equiv (e \sqrt \pi a_\beta / a )^3
$
is the number of particles at which the chemical potential changes its sign.
The energy functional for the logarithmic BEC can be defined
in a standard way (see, e.g., Sec 6.1 of the book \cite{pethick04}):
\be
E [n] =
4 \pi \int\limits_0^\infty
\left\{
\frac{\hbar^2}{2 m n}
(\vec\nabla \sqrt n)^2
%+ \frac{1}{2} m \omega^2 r^2
-
\beta^{-1}
\left[
\ln{(n a^3)} -1
\right]
\right\}
n
r^2 d r
,
\ee
which yields the expression
for the energy per particle:
\be\lb{e-ennotrap}
E / N =
3 \beta^{-1}
\left[
\frac{4}{3}
-
\ln{
\left(
\frac{N^{1/3}}{\sqrt{\pi}}
\frac{a}{a_\beta}
\right)
}
\right]
=
\beta^{-1}
\left[
1-
\ln{
\left(
N/N_0
\right)
}
\right]
%= \mu + \beta^{-1}
.
\ee
One can see that $E$ as a function of $N$
vanishes in the origin and at $e N_0$,
and
has a positive-valued local maximum, i.e.,
it is bounded from above (see Fig. \ref{f:e_mu_n})
\be\lb{e-ennotrapmaxN}
E/N \leqslant
\beta^{-1}
,
\ee
where the value of $N$ at which the maximum occurs
is given by the transcendental equation
$\ln{
\left(
N/N_0
\right)
}
+ N_0/N = 1$.
Notice that for the non-negative values
of $E$ there exists some sort of degeneracy -
one value of energy corresponds to two values of $N$.
This will be discussed below in more detail.
Further, if we treat $E$ as a function of the coupling then
it obviously vanishes when $\beta^{-1}$ approaches zero.
However, it can also vanish at the nonzero value of the coupling
which can be made obvious by
rewriting the expression as
\be\lb{e-ennotrap2}
E / N =
- \frac{3}{2}
\beta^{-1}
\ln{
\left(
\beta^{-1} / \beta_0^{-1}
\right)
}
,
\ee
where
$
\beta_0^{-1}
\equiv
\pi e^{8/3} N^{-2/3} E_a
$
represents the non-zero value of the parameter $\beta$
at which this energy vanishes.
From this expression one
can also see  that the energy is bounded from above
\be\lb{e-ennotrapmax}
E (\beta) \leqslant
%E_\text{max} \equiv
\frac{3}{2}
e^{-1}
N \beta_0^{-1}
=
\frac{3}{2}
\pi e^{5/3} N^{1/3}
E_a
,
\ee
with this local maximum occurring at
$\beta_\text{max}^{-1} = \beta_0^{-1} / e = \pi e^{5/3} N^{-2/3} E_a $.

It is important to note that both the chemical potential (\ref{e-munotrap})  and the ground-state energy
(\ref{e-ennotrap}) depend on the parameter $a$ as $\ln{a_{\beta} /a}$ and as we show below this term can only be  positive as $a_{\beta} > a$.
%The plots for the chemical potential and energy per particle for the logarithmic BEC are given in Figs. %\ref{f:chempot-notrap} and \ref{f:energy-notrap}.

\sscn{Effective potential, stability and critical mass}{sec-ntepst}

For the analysis of the solutions of the generic equation
of the form
\[
%\be\lb{e-becgeneq2}
\left[
%- i \hbar \, \partial_t
- \frac{\hbar^2}{2 m} \vec \nabla^2
+
V_\text{ext} (\vec x,\, t)
+
F(|\Psi|^2)
- \mu
\right]
\Psi
= 0
\]
it is often convenient to introduce the notion
of the effective potential
$
V_\text{eff} (r)
\equiv
V_\text{ext} (\vec x,\, t)
+
F(|\Psi (r)|^2)
$
which in our case is defined as
\be
V_\text{eff} (r)
\equiv
%\frac{1}{2} m \omega^2 r^2
-
\beta^{-1} \ln{(a^3 \Psi^2(r))}
.
\ee
Being evaluated on the solution
(\ref{e-solnotrap})
it becomes the quadratic function
of radius-vector:
\be
V_\text{eff} (r) =
\frac{1}{2}
m
\omega_\text{eff}^2
%\frac{m \chi '^2}{\hbar^2}
r^2
+
\mu
-
\frac{3}{2} \hbar \omega_\text{eff}
,
\ee
where
\be
\omega_\text{eff} =
%\epsilon_{\omega\beta}/ \hbar =
2 \beta^{-1}/\hbar
.
\ee
Therefore, the effect from the logarithmic term here
is that it effectively introduces the trap with the  frequency
which is controlled by $\beta$ whereas the system can be viewed
as a particle of energy $\mu$ whose center of mass is located in this trap.
Then the stability of the solution against a small perturbation
is ensured by the fact that this frequency is non-negative
if $\beta$ is non-negative.

This is also confirmed by the Vakhitov-Kolokolov stability criterion
\cite{vk73}
which in our case is equivalent to the non-positivity of
the derivative of the chemical potential with respect to $N$.
From Eq. (\ref{e-munotrap}) one can derive that this derivative
equals to $- \beta^{-1}/N$ and thus it is non-positive indeed.

There is, however, an additional issue of stability.
From Eqs. (\ref{e-ennotrap}) or (\ref{e-ennotrap2})
it is clear
that for the non-negative values
of $E$
%(which is the binding energy of the condensate)
there
exists some sort of degeneracy -
one value of energy is achieved at two values of $N$ or $\beta$,
respectively.
This indicates that the system spontaneously `chooses''
either of these values at no cost of binding energy, and the so-called
\textit{mechanical instability} occurs \cite{hs96}.
Thus, our condensate is impeccably stable
only
at
\be\lb{e-enstab}
E < 0
,
\ee
 which means
that the number $N$ must be bounded from below
\be\lb{e-notrapNb}
N
>
N_\text{min}
\equiv
e N_0
.
\ee
The physical meaning of the latter  is
that one cannot
create the stable logarithmic condensate with just any amount of initial (bare)
particles - it must be larger than the value
$N_\text{min}$.
In other words, in the presence of certain long-range correlations and
multi-body scattering the initial Bose system must have the mass larger than the critical one,
$m N_\text{min}$, for the Bose condensation to happen.

\sscn{Self-sustainability and droplet formation}{sec-ssabt}

An important thing to notice is that indeed
the density of the free condensate with the positive $\beta $
does not become the uniform one -
in empty space
the condensate rather tends to take the shape of a spherical Gaussian droplet,
with a characteristic size of order $a_\beta $.
Our droplet, however, is different from the ordinary ones.
As long as its density $n (r)$  is described by the Gaussian the
major portion of
mass and energy is localized in a finite volume $\tfrac{1}{6} \pi a_\beta^3$ - similarly to the
liquid droplets in the classical world.
The difference is that the Gaussian vanishes at the infinity only,
therefore, our droplet is essentially inhomogeneous and
has no sharply defined external surface, hence,
its stability is ensured by the nonlinear quantum effects in the bulk rather than
by the surface tension.

Further, from Eqs. (\ref{e-munotrap}) and (\ref{e-ennotrap}) it follows
that for the Gaussian solution the following relation holds
\be
\mu - E/N = - \beta^{-1}
,
\ee
which essentially means that the chemical potential counted
from the  energy per particle
is negative in our case.
A similar feature was observed in
real superfluids, such as $^4$He \cite{woo76},
therefore, models involving the logarithmic nonlinearity
might find an application there.

\sscn{Quantum Shannon entropy and non-thermal temperature}{sec-ntentrp}

Let us consider now the notion of the information entropy introduced in the Appendix.
In the context of the Bose-condensation phenomenon
the entropy $S_\Psi$ is a measure of information
stored in the microscopical configuration of $N$ bosons which
form the droplet where the square of wave function determines the density value
and constant $\Xi$ defines the information entropy reference
frame.
Besides, $S_\Psi$ becomes a thermodynamic-like measure of the quantum spreading of the droplet as a whole.

To start with  $S_\Psi$, it is natural to measure the constant $\Xi$ in the units of $a^3$:
$
\Xi \equiv Z_0^{1/N} a^3
,
$
where $Z_0$ is a dimensionless constant.
Then the entropy (\ref{e-shent})  can be written as
\be\lb{e-notrapentrp0}
k_B^{-1}
S_\Psi
=
k_B^{-1}
S_\Psi^{(0)}
+ \int n \ln{(a^3 n)} d^3 x
,
\ee
where $S_\Psi^{(0)} \equiv k_B \ln{Z_0}$ is the reference entropy.
One can physically justify the identification of an entropy
using the fact that the energy functional is bounded from  below.
The reasoning is the following.
The energy needed to cut the $N$-particle droplet into two equal parts is
$\Delta E=-N \beta^{-1} \ln 2 $, into three parts it is $\Delta E=-N \beta^{-1} \ln3 $ and so on.
Therefore, the maximum binding
energy which is saved in a droplet and can be realized is no more than
$\Delta Q = N \beta^{-1} \ln N $.
So we can use the thermodynamic definition of entropy through
$\Delta Q= T_\Psi \Delta S_\Psi$.
The parts of the energy (\ref{e-ennotrap}) which are proportional to $N$ do
not contribute to $\Delta E$ and can be taken either to
$E^{(0)}$ or to $S_\Psi^{(0)}$ while $S_\Psi \propto N \ln{N}$:
\ba\lb{e-en_ad}
&&
E=
E^{(0)}
- T_\Psi S_\Psi,
\\&&
S_\Psi=k_B N \ln{N},
\\&&
E^{(0)}=N \beta^{-1} (1+\ln{N_0})
,
\lb{e-en_ad1}
\ea
where $T_\Psi = (k_B \beta)^{-1}$ is the associated temperature.
Thus,  we can formally divide the energy into two types - into the entropy and the part which
is linear w.r.t. $N$.
Therefore, the latter is linear w.r.t. density too, while
$E^{(0)}/N$ can be considered as the energy of a free quasi-particle.
Alternatively,
the $N$-linear part can be considered as a reference entropy
$S_\Psi^{(0)}=N  (1+\ln{N_0})$
which
is supposed to be positive.

In general, such definition of entropy can be
deduced for any shape of the logarithmic BEC (e.g., in the trap)
as the integral in Eq. (\ref{e-notrapentrp0})   can be always presented in the form
$N \ln{N} +N f$ where $f$ is an $N$-independent function of  the parameters of a system.
If we require the positiveness of the additive part of the energy
(it means that $f$ is always positive) the entropy part of the energy Eq. (\ref{e-en_ad})
determines its lower limit at  given  $N$.

\sscn{Emergent spatial extent}{sec-emsi}

From Eq. (\ref{e-solnotrap}) one can see
that while the logarithmic condensate is
``made'' of $N$ bare particles
which are assumed to be point-like
the resulting Gaussian droplet acquires a finite size $a_\beta $.
Besides, it is clear that
the particle density of
the logarithmic condensate  is bounded
from above,
$
n \leqslant n_0
$
,
which implies
that the average volume occupied by
a condensate's quasi-particle
is bounded from below
\be
\langle \text{Vol} \rangle
\geqslant
(\sqrt\pi a_\beta)^3/N
.
\ee
Together with the arguments coming from
the analysis of elementary excitations, see the paragraph after
Eq. (\ref{e-forbidp2}),
this means that the effective size of the
quasi-particle can be non-zero at the non-zero $\beta^{-1}$.
Recalling the meaning of $ a_\beta $ as an effective size of
the Gaussian droplet, we see
that the latter can accommodate no more than
$\tilde N_\text{max} $ quasi-particles,
i.e.,
\be\lb{e-notrapnmax}
\tilde N \leqslant
\tilde N_\text{max}
\approx  \tfrac{1}{6} \pi a_\beta^3 / \langle \text{Vol} \rangle
= N / (6 \sqrt\pi)
,
\ee
where tilded values refer to the condensate's quasi-particle modes.
%Therefore, the value $a_\beta / N^{1/3}$ can be approximately interpreted as the  effective size of the quasi-particle.

However, the inequalities of this type are trivial in a sense; they follow entirely
from the normalization condition and thus they are not necessary specific of the logarithmic model.
In our case, the better bound can be derived from physical arguments.
It is clear that the $N$-linear part  (\ref{e-en_ad1})
must be non-negative because this energy can only increase when $N$ does.
Thus, we have to impose $\ln{N_0} \geqslant -1$ which is equivalent to $n_0 a^3 \leqslant e^4 N$
or $a_\beta \geqslant a/ (\pi^{3/2} e^{4/3}) $ (we remind that $a_\beta $
is the mean size of the Gaussian droplet).
As long as
$
n \leqslant n_0
$,
we obtain
\be\lb{e-ineqpe}
n a^3 \leqslant n_0 a^3 \leqslant N e^4
,
\ee
which implies that $a_r /r \leqslant 1$, where
$a_r = a / (N e^4)^{1/3} $
is the quasi-particle effective size and $r$ is the mean interparticle
distance.
%\footnote{Here we do not specify what is the ``quasi-particle'' size, for example it can be the range of an interaction or the  size of a hard core.}
Thus, the smallest volume of the system of $N$ particles
is nonzero but of the order $a^3$.
%In other words, Eq. (\ref{e-ineqpe}) indicates that if $N$ tends to infinity then the system tries to reach the maximally dense packing with the maximum density being given by $n_{max} = 1/a_r^3$.
%For the finite number of particles we have a droplet with the central density $n_0 < n_{max}$.
This is crucially different from what the GP approach would yield -
%the diluted condensates described by the GP approximation
there
no such effect appears, the Gaussian-type droplet can only
form  when the gas is placed into the
external trapping potential of a special kind.
It is interesting that the spatial extent
exists in the worldvolume-formulated theories
\cite{Green:1987sp}
(where it is postulated from the beginning),
and also it is known to arise as the quantum effect
in the quantum mechanics
of point particles
on non-commutative spaces
\cite{Rohwer:2010zq}.
In our case, however, no spatial extent was initially postulated
(the bare particles
were assumed to be point-like, according to the standard quantum-mechanical approach)
and no quantum commutation rules were modified -
instead,
the non-zero size is an emergent quantum phenomenon
due to the peculiar type of correlations between particles'
wave-functions.

\sscn{Effective potential and finite volume}{sec-ep}

\begin{figure}[htbt]
\begin{center}\epsfig{figure=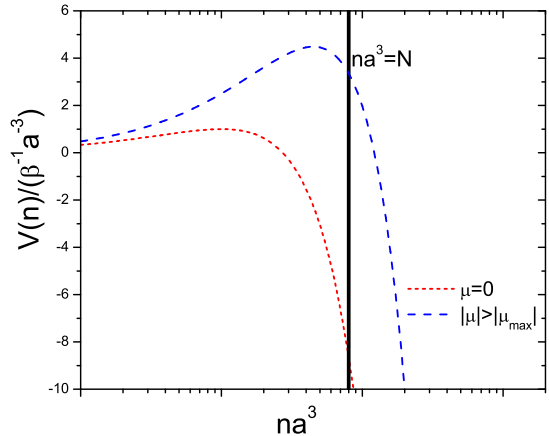,width=  1.05\columnwidth}\end{center}
\caption{
The effective potential (\ref{e-potbm}) for the logarithmic BEC  for zero and large negative chemical potentials}
\label{f:vn}
\end{figure}

As shown above, the value $a^3$ is the smallest possible  volume of the system of $N$ particles
whereas $a_r \sim a/N^{1/3}$ is the size of a quasi-particle (for brevity, in this section
we neglect the ${\cal O} (1)$ factors such as those containing $e$ and $\pi$).
%Then it is natural to require that the average distance between particles is larger than this size, which  means $na^3<N$. If $na^3 \sim N$ then we have a compact packing in our system.
One can argue that
the condition (\ref{e-ineqpe})
means that states with  $n a^3 > N$
 can not be physically reached as particles commence to ``overlap'' each other.
Therefore,
the effective potential,
\be\lb{e-potbm}
V (n) \equiv V_\beta - \mu n
,
\ee
is confined to the given bounds, see Fig. \ref{f:vn}.
In particular,
it is naturally bounded from  below because $n a^3 \ln{(n a^3)} \ll N \ln{N}$.
A similar condition holds for the central density which implies that $a_{\beta}/a > 1$
(the droplet's volume  must  not be smaller than the smallest one).
Therefore, our potential for the positive $\beta$ has two minima,
one being located at zero, and the other being at the limit value
corresponding to maximally dense packing.
The latter is usually deeper and hence defines the ground state.
For the system with a finite number of particles, such as our droplet,
the density of a dense packing is much larger than its central density $n_0$.
The depth of this minimum can be controlled with a quadratic
term, see Fig. \ref{f:vn}.
If one considers this potential with the extra quadratic (chemical potential) term one can find that
negative chemical potentials shift the local maximum to the large values. At
$\ln{(n a^3)}=1-\mu_{max}/\beta^{-1}$
the minimum at zero becomes deeper and the maximally dense packing state is not the ground state anymore
hence the system can not be self-sustained anymore.

The size of the droplet is defined by a parameter $\beta$ and by putting more and more particles into the system
the maximum density can be reached when $n a^3 \approx N$ which means $a_\beta \approx a$.
One might expect that at a given $N$ one can get a denser condensate  by
putting it inside a trap, but the size of this trap should be smaller than $a_{\beta}$.
In this case the zero minimum of the effective potential is responsible for the ground state.
It should be noted here that if one selects
the harmonic trap to identify such a phase transition
it might be rather difficult as the ground state for both phases is still described
by the Gaussian functions, only of a different width.
A clearer way would be
to use a non-harmonic trap so that a phase transition should reveal itself
through the change in the shape of the droplet.

As for the case of the negative $\beta$, the ground state is determined
there
by the minimum of the effective potential at $n a^3=1$.
We found
that this state can not be self-bound  and can  be confined only in a trap.

\scn{Logarithmic BEC in a trap}{sec-bectrap}

\begin{figure}[htbt]
\begin{center}\epsfig{figure=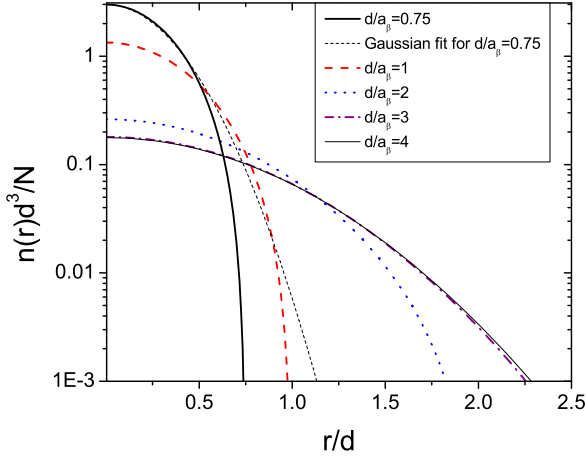,width=  1.05\columnwidth}\end{center}
\caption{
The density of the logarithmic BEC  in the infinite-walled spherical well of radius $d$.}
\label{f:dens}
\end{figure}

First we consider the logarithmic BEC placed in the infinite-walled spherical
square-well potential of radius $d$.
As one can see from (\ref{e-soltrapnorm}), the condensate density in a trap can be much larger
than for a free droplet, see Fig. \ref{f:dens} for the case of positive $\beta$.
When the size of trap $d$ is much larger than the characteristic size of the droplet $a_{\beta}$, the condensate does not
``sense'' the trap and behaves like a self-bounded Gaussian droplet.
As long as $d \sim  a_{\beta}$ the condensate begins loosing its Gaussian shape and becomes denser in the center.
In the case of the negative $\beta$ there is no localized solution and condensate tends to fill the whole trap of any size.
It should be noted that the condensate density profile does not depend on the quasi-particle's size
and only determines the possible number of particles in the condensate.

Now, let us place the logarithmic condensate into the isotropic harmonic trap
of frequency $\omega$.
Then the equation (\ref{e-becgeneq}) reduces to
the following one
\be\lb{e-odetrap}
- \frac{\hbar^2}{2 m}
\Delta \Psi
+
\left[
\tfrac{1}{2}
m \omega^2 r^2
- \mu
-
\beta^{-1} \ln{(a^3 \Psi^2)}
\right]
\Psi
= 0,
\ee
where
$a_\omega = \sqrt{\hbar / m \omega}$
is the characteristic width of the harmonic trap.
Unlike the previous (free) case,
the exact normalized solution of this equation
exists both for positive and negative $\beta $'s
and still has the Gaussian form:
\be\lb{e-soltrapnorm}
n(r) =
n_0 (\omega)
%\exp{\left(-\frac{1-\chi}{2 A_\beta} r^2  \right)}
e^{-(r/a_{\beta\omega})^2}
,
\ee
where
\be
n_0 (\omega)
\equiv
%N \left(\frac{1-\chi}{2 \pi A_\beta}\right)^{3/2}
\pi^{-3/2} N/ a_{\beta\omega}^3
,
\quad
a_{\beta\omega}
\equiv
\sqrt{\frac{2 A_\beta}{1-\chi}}
=
\frac{\hbar}{\sqrt{m \epsilon_{\omega\beta}}}
,
\ee
and
\be
\chi \equiv - \Sign{\beta} \sqrt{1 + (2 A_\beta/a_\omega^2)^2}
=
- \Sign{\beta} \sqrt{1 + (\hbar \omega \beta)^2}
,
\ee
so that the values $\epsilon_{\omega\beta} = (1-\chi)/\beta$
and $\sigma$ are never negative.
%here we also denoted $a_\mu = \hbar / \sqrt{ 2 m \mu }$.
The normalization condition yields
the allowed value of the chemical potential:
\bw
\be\lb{e-mutrap}
\mu_\omega =
\frac{3}{2} \beta^{-1}
\left[
1-\chi -
\ln{
\left(
\frac{ m a^2 N^{2/3}}{\pi \hbar^2}
\frac{1-\chi}{\beta}
\right)
}
\right]
=
-
\beta^{-1}
\ln{
\left(
N / N_0 (\omega)
\right)
}
,
\ee
\ew
where
\be
N_0 (\omega)
\equiv
\pi^{3/2}
%\left(\frac{\hbar}{a}\right)^3
\left(
%\frac{\pi}{m \epsilon_{\omega\beta}}
a_{\beta\omega}/a
\right)^{3}
e^{
%\left[
\frac{3}{2} (1 - \chi)
%\right]
}
,
\ee
and
in the limit $\omega \to 0$ at the positive $\beta$
the expressions expectedly reduce to those from the
previous section, $N_0 (\omega = 0) = N_0$.

By analogy with the spherical-wall case one can
show that when  $\hbar \omega \gg \beta^{-1}$
(which is equivalent to $a_\omega < a_\beta$) the central
density in the trap can be considerably larger than the central density of the droplet.
But of course it can not be infinitely large because the condensate density has an upper
bound (\ref{e-ineqpe}).

\begin{figure}[htbt]
\begin{center}\epsfig{figure=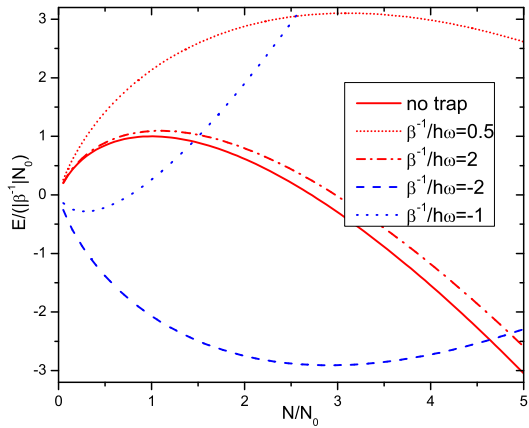,width=  1.05\columnwidth}\end{center}
\caption{
The energy of the logarithmic BEC in an isotropic harmonic trap versus the number of particles
for different values of $\beta^{-1}/\hbar \omega$.
}
\label{f:e_w}
\end{figure}

Further, the energy of the system is given
by
\bw
\be
E [n] =
4 \pi \int\limits_0^\infty
\left\{
\frac{\hbar^2}{2 m n}
(\vec\nabla \sqrt n)^2
+
\frac{1}{2}
m \omega^2 r^2
-
\beta^{-1}
\left[
\ln{(n a^3)} -1
\right]
\right\}
n
r^2 d r
,
\ee
hence the energy per particle
becomes
\be\lb{e-entrap}
E_\omega / N =
\frac{3}{2} \beta^{-1}
\left[
\frac{5}{3} -  \chi
-
\ln{
\left(
\frac{m a^2 N^{2/3}}{\pi \hbar^2}
\frac{1-\chi}{ \beta}
\right)
}
\right]
=
\beta^{-1}
\left[
1-
\ln{
\left(
N/N_0(\omega)
\right)
}
\right]
.
\ee
\ew
The relation
$\mu_\omega - E_\omega / N = - \beta^{-1}$ holds, similarly to the case
of the condensate without a trap,
which
means that the chemical potential counted
from the condensate energy per particle
is negative (positive) for the positive (negative) $\beta$.
The plot of the energy in  $\beta^{-1} N_{0}$ units
for the logarithmic BEC in an isotropic harmonic trap
as functions of a number of particles in $N_{0}$ units
is given in Fig. \ref{f:e_w}.
% For the positive $\beta$ the energy

\sscn{Effective potential and linear stability}{sec-epst}

The effective potential in this case is defined as
\be
V_\text{eff} (r)
\equiv
\frac{1}{2}
m \omega^2 r^2
-
\beta^{-1} \ln{(a^3 \Psi^2(r))}
.
\ee
Being evaluated on the solution
(\ref{e-soltrapnorm})
it becomes:
\be
V_\text{eff} (r) =
\frac{1}{2}
m
\omega_\text{eff}^2
%\frac{m \chi '^2}{\hbar^2}
r^2
+
\mu_\omega -
\frac{3}{2} \hbar \omega_\text{eff}
,
\ee
where
\be
\omega_\text{eff} = \epsilon_{\omega\beta}/ \hbar =
\beta^{-1}
\left(
1 + \Sign{\beta} \sqrt{1 + (\hbar \omega \beta)^2}
\right)/\hbar
.
\ee
Therefore, the effect from the logarithmic term here
is that it replaces the trap frequency
with the modified one which depends on $\beta$
and it is always non-negative regardless of the signs of $\omega$
or $\beta$.\footnote{This effect is similar to what happens in quantum mechanics on non-commutative spaces where the motion of the particle placed in
a harmonic trap stays oscillatory but it is determined not by the bare frequency but by
the modified one containing the non-commutativity parameter \cite{Rohwer:2010zq}.}
Thus, the coefficient at $r^2$ is non-negative
regardless of the sign of $\beta$ either, hence,
the Gaussian solution (\ref{e-soltrapnorm}) is stable both at a positive
and negative $\beta$.
Moreover, it stays stable at arbitrary small $\hbar \omega\beta$
unless the latter vanishes exactly - then the negative-$\beta$ solution decays,
in agreement with the result from the previous section.
This feature shows another difference of the logarithmic BEC from
the GP one - the latter decays when the non-linear coupling (proportional
to the scattering length) becomes negative.
These properties are essentially rooted in the sign-changing property of the logarithm
discussed in the introductory part of the paper.

\sscn{Mechanical stability and related bounds}{sec-mst}

As for the mechanical stability the arguments given
for the no-trap case can be easily generalized for our case:
for the logarithmic condensate
to be stable in a harmonic trap the following inequality
must hold
\be\lb{e-negatE}
\beta^{-1}
\left[
1-
\ln{
\left(
N/N_0(\omega)
\right)
}
\right]
< 0
%\beta^{-1} (1+\ln{N_0 (\omega)}) > 0
,
\ee
whose physical implications are to be discussed now.

%\sscn{Frequency-dependent bounds for density and $N$}{sec-emsi2}
%Let us consider the main stability condition (\ref{e-negatE}) and associated bounds.

As long as the coupling is not necessarily positive
now we must distinguish two cases:

(a) Positive $\beta$.
The inequality (\ref{e-negatE})
brings, as in the no-trap case, the lower bound for the number of
condensate particles,
with the only difference that now
it becomes $\omega$-dependent
\ba
&&
N >
N^{(+)} =
e N_0 (\omega)
=
\pi^{3/2}
\left(
\frac{a_{\beta\omega} }{a}
\right)^3
e^{\tfrac{1}{2} (3 |\chi| + 5)}
\nn \\&&\qquad
=
\left(
\frac{2 \pi}{|\chi| + 1}
\right)^{3/2}
\left(
\frac{a_\beta }{a}
\right)^3
e^{\tfrac{1}{2} (3 |\chi| + 5)}
,
\ea
so it can be regulated by adjusting the trap's frequency.
Similarly to the no-trap case, this bound implies that one can not
create the logarithmic condensate with just any amount of initial (bare)
particles - it must be larger than the critical value
$N^{(+)}$ which is frequency-dependent now.
If the frequency energy is much smaller than the logarithmic coupling,
$\hbar \omega \ll \beta^{-1}$,
then the critical value is a quadratic function of the frequency:
\be
N^{(+)}
=
e N_0
\left[
1 + \frac{3}{8}
(\hbar \omega \beta)^2
\right]
+ {\cal O} \left( (\hbar \omega \beta)^{4} \right)
,
\ee
where $N_0$ was defined in the no-trap case.
If the frequency energy is much larger than the logarithmic coupling,
$\hbar \omega \gg \beta^{-1}$,
then the critical value grows exponentially with an increasing frequency:
\be
N^{(+)} \approx  \pi^{3/2} (a_{\omega}/ a)^3
\exp{
\left(
\tfrac{3}{2} \hbar \omega \beta
\right)
}
%+ {\cal O} \left( (\hbar \omega \beta)^{-1} \right)
,
\ee
which means that the trapping potential dominates over the logarithmic term in the wave equation,
and the condensate goes into another phase.

(b) Negative $\beta$.
There appears the upper bound for the number of
condensate particles
which is also $\omega$-dependent
\ba
&&
N
<
N^{(-)} =
e N_0 (\omega)
=
\pi^{3/2}
\left(
\frac{a_{\beta\omega} }{a}
\right)^3
e^{-\tfrac{1}{2} (3 |\chi|- 5)}
\nn\\&&\qquad
=
\left(
\frac{2 \pi }{|\chi|-1}
\right)^{3/2}
\left(
\frac{ \sqrt{|A_\beta|} }{a}
\right)^3
e^{-\tfrac{1}{2} (3 |\chi|- 5)}
.
\lb{e-trapNm}
\ea
This critical value can be also regulated by the trap's frequency.
If the frequency energy is much smaller than the logarithmic coupling,
$\hbar \omega \ll \beta^{-1}$,
then the critical value scales as the inverse cube of the frequency:
\be
N^{(-)} \approx e \pi^{3/2} (a_{\beta\omega m}/ a)^3
%+ {\cal O} (\hbar \omega \beta)
,
\ee
where $a_{\beta\omega m} \equiv a_\omega^2/\sqrt{|A_\beta|} = \omega^{-1} \sqrt{\frac{2}{m |\beta|}} $
is another length scale.
When the frequency vanishes, $N^{(-)}$ diverges which indicates that
at negative $\beta$ there is no
Gaussian solution for the logarithmic condensate in empty space,
as was mentioned in the previous section.
If the frequency energy is much larger than the logarithmic coupling,
$\hbar \omega \gg \beta^{-1}$,
then the critical value vanishes exponentially with increasing frequency:
\be
N^{(-)} \approx  \pi^{3/2} (a_{\omega}/ a)^3
\exp{
\left(
- \tfrac{3}{2} \hbar \omega |\beta|
\right)
}
%+ {\cal O} \left( (\hbar \omega \beta)^{-1} \right)
,
\ee
as the trapping potential dumps the logarithmic-term effects
and thus destroys the logarithmic condensate.

\scn{Excitations in the logarithmic BEC}{sec-exc}

Small fluctuations in quantum Bose liquids can be conditionally
cast into two classes - collective modes and elementary
excitations.
The former describe the motion of the liquid
or a large part thereof as a whole, the latter are localized
and can be viewed as particle-like objects.
Although, due to the particle-wave duality
there is no exact border between these classes: for instance,
a phonon which is a quantum of collective vibrational modes
can be also described as an elementary excitation via the Bogoliubov
approach.

The collective mode of the logarithmic BEC is
governed by the second-rank tensor (``acoustic metric'')
corresponding to the conformally-flat Lorentzian manifold \cite{Zloshchastiev:2009aw}.
That result can be easily transferred to the condensed matter physics
provided that the fundamental velocity constant is assumed
to be the speed of sound.
Here we consider the other class, elementary excitations,
for
the logarithmic condensate in the empty space.
To this effect we perform the second
quantization by going from the wave-functions $\Psi $
to the field operators
via the Bogoliubov decomposition
\be
\Psi \to
\hat\Psi = \sqrt{n_0} + \delta \hat\Psi
,
\ee
where
$n_0$ is the density of the condensate (for simplicity
we assume the latter being uniform
which is a valid approximation at the length scales are smaller compared
with the size of the droplet),
$\delta \hat\Psi $ is
the operator of excitations,
$
[\delta \hat\Psi , \delta \hat\Psi^\dagger] = \theta
$,
and throughout the paper the non-polynomial functions of operators
are defined as the Taylor series of matrices, as usual.
One of the simplest forms of the Hamiltonian
can be easily deduced from Eq. (\ref{e-xmain})
assuming the normal-ordering approximation
\be
\hat H =
\int d {\bf r}
\left[
-
\hat\Psi^\dagger
\frac{\hbar^2}{2 m}
\Delta \hat\Psi
-
\beta^{-1}
:
\hat\Psi^\dagger
\hat\Psi
\ln{(a^3 \hat\Psi^\dagger \hat\Psi )}
:
\right]
%+ {\cal O} (\theta^2)
,
\ee
where the colon means the normal ordering of the operators.
Upon applying the Bogoliubov decomposition, we obtain
\ba
&&
\hat H - \mu \hat N =
\left[
\epsilon_0
-
\beta^{-1}
\left(
1 + \ln{(a^3 n_0)}
\right)
\right]
\delta \hat\Psi^{\dagger}
\delta \hat\Psi
\nn\\&&\qquad\qquad\quad
-
\frac{\beta^{-1}}{2}
\left(
\delta \hat\Psi^2
+
\delta \hat\Psi^{\dagger 2}
\right)
,
\ea
when omitting non-quadratic terms;
here we denote
$
\epsilon_0 \equiv p^2 / 2 m - \mu
$,
$p $ being the absolute value of the momentum vector.
Upon applying the Bogoliubov diagonalization procedure we obtain
\ba
&&
\hat H - \mu \hat N
=
\epsilon \hat b^\dagger b
,
\\&&
\epsilon^2
\equiv
\left[
\epsilon_0
-
\beta^{-1}
\left(
1 + \ln{(a^3 n_0)}
%+ g_1 (n_0)
\right)
\right]^2
-
\beta^{-2}
%\left[1 + g_2 (n_0)\right]^2
%\right\}^{1/2}
,
\ea
where
$\hat b^\dagger$ and $\hat b$
are the creation and annihilation operators
of the quasi-particles with momenta $\textbf{p}$,
$[\hat b, \hat b^\dagger]
=
[\delta \hat\Psi , \delta \hat\Psi^\dagger]
$,
defined as the linear combinations
\ba
&&
\delta \hat\Psi =
\kappa \hat b + \sqrt{\kappa^2 -1} \, \hat b^\dagger
,
\nn\\&&
\delta \hat\Psi^\dagger =
\sqrt{\kappa^2 -1}\, \hat b + \kappa \hat b^\dagger
,
\nn
\ea
where
\ba
&&
\kappa
\equiv
- \frac{\kappa_2}{\sqrt 2}
\left(
\kappa_1^2 - \kappa_2^2
\right)^{-1/4}
\left(
\kappa_1
-
\sqrt{
\kappa_1^2 - \kappa_2^2
}
\right)^{-1/2}
,
\nn\\&&
\kappa_1
\equiv
\epsilon_0
-
\beta^{-1}
\left(
1 + \ln{(a^3 n_0)}
%+ g_1 (n_0)
\right)
,
\nn\\&&
\kappa_2
\equiv
\beta^{-1}
%\left[1 + g_2 (n_0)\right]
,
\nn
\ea
%Omitting the chemical-potential term,
and we can write the dispersion relation
for the excitations of the logarithmic BEC
in the form
\be\lb{e-disp1}
\epsilon
(p)
=
\sqrt{
\left[
\frac{p^2}{2 m}
-\mu
-
\beta^{-1}
\left(
1 + \ln{(a^3 n_0)}
%+ g_1 (n_0)
\right)
\right]^2
-
\beta^{-2}
%\left[ 1 + g_2 (n_0)\right]^2
%\right\}^{1/2}
}
,
\ee
and also we assume $\beta $ to be positive
for the remainder of this section\footnote{The condensate with the negative $\beta $ in the empty space
can not be described by the Gaussian, therefore, its treatment would be more complex and model-dependent.
Although, one cannot exclude that some results of this section will also be applicable  for
the negative-$\beta $ sector of the model.},
therefore, the value
of the chemical potential can be borrowed
from Eq. (\ref{e-munotrap}).
As compared with the dispersion
relation for the excitations in the GP condensate \cite{pethick04},
\[
\epsilon_\text{GP}
(p)
=
\sqrt{
\frac{p^2}{2 m}
%-\mu\right)
\left(
\frac{p^2}{2 m}
+
2
n_0 U_0
\right)
}
,
\]
the dispersion for the logarithmic Bose liquid
has a number of drastic differences.

The first thing to notice is that
the requirement of non-negative energy
points out  the existence of the forbidden
region of momenta (where $\epsilon$ becomes negative
or even complex which indicates instability issues).
The forbidden momenta region is given by the inequality
(we are neglecting the chemical potential for the moment)
\bw
\ba
\beta^{-1}
%\left[
\ln{(a^3 n_0)}
%+ 1\right]
\leqslant
\left(
\frac{p^2}{2 m}
\right)_\text{forbid}
\leqslant
\beta^{-1}
\left[
\ln{(a^3 n_0)}
+ 2
\right]
\quad
&
\text{if}
&
\quad \ln{(a^3 n_0)} > 0
,
\lb{e-forbidp1}\\
0
\leqslant
\left(
\frac{p^2}{2 m}
\right)_\text{forbid}
\leqslant
\beta^{-1}
\left[
\ln{(a^3 n_0)}
+ 2
\right]
\quad
&
\text{if}
&
\quad \ln{(a^3 n_0)} \leqslant 0
,
\lb{e-forbidp2}
\ea
\ew
and this region only exists if
the upper bound
$\ln{(a^3 n_0)} + 2$ is positive-valued.
The latter condition is equivalent
to $r_0/e^{2/3} > a $ where $r_0 = n_0^{-1/3}$
is the average spacing between quasi-particles.
Therefore, the appearance of the gap in
the momentum space
is in fact
another manifestation of the emergent spatial
extent phenomenon discussed above.

Thus, depending on the value of the background condensate density
the momentum space of the model can contain forbidden regions
and thus can have a non-trivial topology.
To see the full picture, we introduce
two auxiliary values of dimensionality energy
\ba
&&
%p \geqslant
\epsilon_{1} \equiv
\beta^{-1}
%\left(
\ln{(a^3 n_0)}
%+ 1\right)
+ \mu
,
%\quad
\\&&
%\epsilon (p_{0}) = 0,
\epsilon_{2} \equiv \epsilon_{1} + 2 \beta^{-1}
> \epsilon_{1}
,
\ea
then
the dispersion relation (\ref{e-disp1})
can be rewritten in the form
\be\lb{e-disp2}
\epsilon
(p)
=
%\frac{1}{2 m}
\sqrt{
\left(
\frac{p^2}{2m} - \epsilon_{1}
\right)
\left(
\frac{p^2}{2m} - \epsilon_{2}
\right)
}
,
\ee
which is more convenient when classifying dispersion
relations according to the topology of the momentum space.
We distinguish the following cases:

(i) $\epsilon_2 < 0$. Then $\epsilon_1$ must be negative as well, therefore,
energy vanishes nowhere and
the topological structure of the momentum space is trivial.
The quasiparticle's spectrum is fully gapped and equivalent to that of
the massive relativistic particle moving in the 4D spacetime with the fundamental velocity
constant determined by the speed of sound $\bar c$:
\be
\epsilon
=
\sqrt{
p^2
\bar c^2
+
m_\star^2 \bar c^4
}
+
{\cal O} (p^4/m^2)
=
\Delta +
\frac{p^2}{2 m_\star}
+
{\cal O} (p^4/m^2)
,
\ee
where
\ba
&&
\Delta = \sqrt{\epsilon_1  \epsilon_2},
\ \
m_\star = - \frac{2 \Delta}{\epsilon_1  + \epsilon_2} m,
\nn\\&&
\bar c = \sqrt{- \frac{\epsilon_1  + \epsilon_2}{2 m}}
=
\sqrt{\frac{\beta^{-1}  - \epsilon_2}{m}}
.
\ea
To allow the excitations of this type,
the background condensate density
must be below the first threshold value
\be
n_0 <
\ntr
,
\ee
where
\be
\ntr \equiv
a^{-3}
e^{-\mu \beta -2}
=
\frac{1}{e^{2}
a^{3}}
\frac{N}{N_0}
=
\frac{N}{e^5 \pi^{3/2} a_\beta^3}
.
\ee

(ii) $\epsilon_2 = 0$. Then $\epsilon_1 = - 2 \beta^{-1}$ and thus it is still negative,
in the momentum space
there appears a zero-energy point,
at $p = 0 $, so the topological structure is still trivial and this case
can be treated as a limit of the previous one.
The quasi-particle's spectrum in the vicinity of this point
%(where the energy minimum is reached)
is equivalent to that of a phonon
or a
massless relativistic
particle moving in a spacetime with the fundamental velocity
constant $c_0$:
\be
\epsilon^2 =
c_0^2 p^2
+
{\cal O} (p^4)
,
\ee
where the speed of sound is given by
\be
c_0 \equiv 1/\sqrt{m \beta}
,
\ee
so that $c_0 =  c_s$,
cf. Eq. (\ref{e-ssnd}).
To allow the excitations of this type,
the background condensate density
must be equal to the first threshold
\be
n_0 =
\ntr
.
\ee

(iii) $\epsilon_2 > 0$, $\epsilon_1 < 0$.
To allow the excitations of this type,
the background condensate density
must lie between the first and second threshold values
\be
\ntr
<
n_0
<
e^2
\ntr
,
\ee
then
in the momentum space
there appears a zero-energy sphere of radius
$p = \sqrt{2 m \epsilon_2}$.
The positive-energy modes are only located
 outside it, the quasiparticle's spectrum in the vicinity of this sphere
is equivalent to that of a
tachyonic particle moving in a spacetime with the fundamental velocity
constant $c_2$:
\be\lb{e-disptachyon}
\epsilon^2 =
p^2 c_2^2
+
m_2^2 c_2^4
+
{\cal O} \left((p^2 - 2 m \epsilon_2)^2 \right)
,
\ee
where the speed of sound and imaginary mass are
\be
c_2 = \sqrt{ \frac{\epsilon_2 - \epsilon_1}{2 m}},
\ \
m_2 = 2 i \sqrt{\frac{\epsilon_2}{\epsilon_2  - \epsilon_1}} m
.
\ee
In particle physics such models are usually rejected as incompatible
with Lorentz symmetry but in our case this can not be a valid reason:
this symmetry is not a fundamental 
but rather an emergent phenomenon which appears at the lowest levels of energy.

(iv) $\epsilon_2 > 0$, $\epsilon_1 = 0$.
This happens when the background density approaches the second threshold
%(where the logarithm changes sign)
\be
n_0 =
e^2
\ntr
,
\ee
then
in the momentum space
there appears the zero-energy sphere of radius
$p = \sqrt{2 m \epsilon_2}$
and the zero-energy point at $p=0$.
This leads to degeneracy and the appearance of two zero-energy states.
The first mode is identical to the mode described in the previous
item upon setting $\epsilon_1$ to zero.
The other mode, the one with vanishing $p$,
has the formal dispersion relation:
\be
\epsilon^2 =
- \frac{\epsilon_2}{2 m} p^2
+
{\cal O} \left( p^4 \right)
,
\ee
which is of the quaternionic or 4D Euclidean ($SO(4)$-symmetric) type.
However, this dispersion is only valid  at $p = 0$, therefore,
this mode is non-propagating.

(v) $\epsilon_2 > 0$, $\epsilon_1 > 0$.
This corresponds to the density growing above the second threshold
\be
n_0 >
e^2
\ntr
,
\ee
then
the momentum space gets
split by two zero-energy spheres, of radii
$\sqrt{2 m \epsilon_1}$ and $\sqrt{2 m \epsilon_2}$,
%($\epsilon_2 - \epsilon_1$ is always positive),
which leads to degeneracy and the appearance of two zero modes.
Positive-energy modes form two bands, $p < \sqrt{2 m \epsilon_1}$ and $p > \sqrt{2 m \epsilon_2}$.
The first zero mode, located near the internal sphere $\sqrt{2 m \epsilon_1}$, is
described by the dispersion:
\be
\epsilon^2 =
\frac{\epsilon_1 - \epsilon_2}{m} p^2 +
2
\epsilon_1 (\epsilon_2 - \epsilon_1)
+
{\cal O} \left((p^2 - 2 m \epsilon_1)^2 \right)
,
\ee
so
it belongs to the above-mentioned quaternionic type with the only
difference that now
its energy is positive and the momentum does not vanish.
The other zero-energy mode is of the tachyonic type, it is equivalent
to the one described by Eq. (\ref{e-disptachyon}).

To conclude, the topology of the momentum space
of the logarithmic BEC excitations depends on the background value
of the condensate's density.
In its turn,
depending on the topology of the momentum space their dispersion relations can cover all cases
described by the symmetries of the special orthogonal groups of order four -
massive relativistic,
massless relativistic, tachyonic and quaternionic.
The tachyonic (quaternionic) mode arises when the momentum space contains the zero-energy
sphere
so that positive-energy states are located outside (inside) it.
One can also notice that the dispersion relations become more
and more exotic as the background density grows.
This is in fact another manifestation of the above-mentioned
emergent size effect.
%in the logarithmic BEC it is not possible to make the interparticle distance vanish exactly.

\scn{Conclusion}{sec-con}

In this paper we considered the toy model based on the
logarithmic \schrod equation and
applied it to describing the ground state and
excitations of some hypothetical Bose condensate.
That is why we did not specify why and how such sort of nonlinearity
might exist in real Bose liquids.
We only pointed out that LogSE,  despite
its simplicity, reveals an interesting class of solutions and it
can  possibly be adapted for a description of strongly- interacting Bose liquids.
The important feature of the logarithmic BEC
is that it can exist as a self-bound  droplet since the logarithmic nonlinearity
uniquely combines both an attractive and a repulsive interaction.
But a self- sustainable solution may exist for the generalized GP equation in the case of a very large scattering length as well, where the combination of an attractive two-body interaction and a repulsive three-body interaction
allows to manufacture self-bound quantum droplets \cite{Bulgac2002}.
Though one may suppose that such systems with a large scattering length are not ``safe'' against
$N$-body interactions and, for example, a four-body interaction might destroy a droplet
if it is too attractive while some high-order interaction might stabilize it.
However, the logarithmic type of an interaction naturally
has both  the attractive and the repulsive parts which are formally neither
two-body nor any other $N$-body interactions.
Therefore,
LogBEC  does not need to introduce any extra interaction terms in order to get a stable self-sustainable solution.

We found that the self-bound ground state, Gaussian droplet, can only exist at a positive parameter $\beta$ and its energy is bounded both from below and from above. It is provided by the fact that the potential energy can be both positive and negative whose behaviour is defined by its logarithmic dependence on density.
The effective potential for such state is of the harmonic trap type whose frequency is defined by a parameter $\beta$.
The logarithmic BEC with a negative parameter $\beta$ can
not have the self-bound state and only exists  in a trap.
We found that in the harmonic trap the solution is always of the Gaussian type for the parameter $\beta$ of any sign and value.

We justified the existence of the characteristic quasi-particle's size  which
limits the maximum density and the ground state energy of the condensate.
This characteristic size naturally arises if we directly require
the positiveness of the additive part of the ground state energy (\ref{e-en_ad}).

Finally, we studied the elementary excitations  and demonstrated that
depending on the topological structure of the momentum space
their dispersion relations can be of the massive relativistic,
massless relativistic, tachyonic and quaternionic type.

\begin{acknowledgments}
K.Z. is grateful to D. Anchishkin, D. Churochkin and Yu. Sitenko
for fruitful discussions, as well as to A. A. for supporting his
visits to the
Stellenbosch node of the National Institute for Theoretical Physics.
This work was supported under a grant of the National Research Foundation of South Africa.
\end{acknowledgments}

\appendix*
\scn{Logarithmic \schrod Equation}{sec-app}

There exist at least two ways of how
the logarithmic \schrod equation (LogSE) can be
introduced.
The original one is based on the separability
argument - the LogSE is the only local \schrod equation
(apart from the conventional linear one)
which preserves the separability
of the product states:
the solution of the LogSE for a composite system is a product
of the solutions for  uncorrelated subsystems \cite{BialynickiBirula:1976zp}.

The second way is based on arguments which come
from open quantum systems and quantum information
theory \cite{Yasue:1978bx}.
It is relatively
less known and thus deserves to be mentioned here.
Let us consider a multi-particle (sub)system
whose dynamics is described by the Hamiltonian-type
operator $\hat{{\bf H}}$.
Besides, this subsystem is in  contact
with its environment so that there is an exchange
of energy and information.
The state of the system is described by the
vector $|\Psi\rangle$.
If the Hamiltonian
does not depend on a wave function then
in the \schrod coordinate representation
we recover the linear differential equation for $\Psi$.

However, in general the interactions between the particles
comprising the subsystem depend on the distribution $|\Psi|^2$
of the particles in the configuration space.
To determine this distribution, i.e., to extract, transfer and store the information
in a particular configuration of matter,
one requires a certain amount of energy per bit, call it
$\varepsilon $.
The information acquired upon measurement of the state
is proportional to the logarithm of the probability
of an outcome $\Psi$, i.e.,
\be
I_\Psi = -\log_2 (\Xi |\Psi|^2) = -  \ln (\Xi |\Psi|^2)/ \ln 2,
\ee
and the associated Shannon entropy of the subsystem
is given by\footnote{As far as we know,
this kind of entropy was introduced first in Ref. \cite{ever55} and subsequently
rediscovered by several authors afterwards.}
\be\lb{e-shent}
S_\Psi = - \zeta k_B \langle \Psi | \ln{(\Xi |\Psi|^2)}|\Psi \rangle
,
\ee
where $k_B$ is the Boltzmann constant,
$\zeta $ is the sign function chosen in such a way that $S_\Psi$ stays
positive.
This entropy approaches minimum on delta-like distributions and maximizes on uniform
ones, therefore, it can be used as a measure of
``spreading'' of the probability distribution described by $\Psi$.
Here the normalization factor $\Xi $ defines a measurement
reference for the entropy because for continuous systems
the latter is not absolute.
For instance, one could establish the reference entropy as that for a uniform
distribution hence
if the subsystem has a fixed volume and the states are box-normalized
then $\Xi$ equals this volume.

It should be also mentioned that
this entropy is closely related to the logarithmic Sobolev inequality
(known by physicists as the Everett-Hirschmann uncertainty relation)
which was conjectured by Everett \cite{ever55},
rediscovered by Hirschmann \cite{hir57} and
proven by Beckner \cite{beck75} using the Babenko inequality \cite{bab61}.
It has been shown that this uncertainty relation is stronger
than the Heisenberg one, among its other applications
is the proof that the energy in logarithmic nonlinear quantum
mechanics is bounded from below \cite{BialynickiBirula:1976zp}.
Essentially,
the Everett-Hirschmann uncertainty means the following:
suppose we have the probability densities in the
$d$-dimensional space and its Fourier transform
which are normalized to $N$,
$
\int  |\Psi (\textbf{r})|^2 d^d \textbf{r} =
\int |\Psi (\textbf{k})|^2  d^d \textbf{k}  = N
$,
$\Psi (\textbf{k}) = {\cal F} \Psi (\textbf{x})$.
Then the Everett-Hirschmann inequality reads
\ba
&&
- \int  |\Psi (\textbf{r})|^2 \ln{|\Psi (\textbf{r})|^2} d^d \textbf{r}
- \int  |\Psi (\textbf{k})|^2 \ln{|\Psi (\textbf{k})|^2} d^d \textbf{k}
\nn\\&&\quad
\geqslant
d (1+\ln\pi) N
-2 N \ln N
,
\lb{e-ehin}
\ea
where the l.h.s terms are easily recognizable as
$S_\Psi$ for the position and momentum space, respectively.
To prove that this inequality is stronger
than the Heisenberg uncertainty,
one has to use Eq. (\ref{e-ehin}) to show that
\ba
&&
- \int  |\Psi (\textbf{r})|^2 \ln{|\Psi (\textbf{r})|^2} d^d \textbf{r}
%- \int  |\Psi (\textbf{k})|^2 \ln{|\Psi (\textbf{k})|^2} d^d \textbf{k}
\nn\\&&
\leqslant
\frac{d}{2}
N
\left\{
\ln{
\left[
\frac{2 \pi e }{d}
(\textbf{r} - \langle \textbf{r} \rangle)^2
\right]
}
-
(1 + 2/d)
\ln N
\right\}
,
\qquad
\ea
and
similarly for $\Psi (\textbf{k}) $;
here $ \langle \ . \ \rangle $
%= \int  \Psi^* (\textbf{r}) \ . \ \Psi (\textbf{r}) d^d \textbf{r}$,
means an average,
as  usual.

Let us go back to the Hamiltonian.
The above-mentioned energy $\varepsilon$ thus contributes
to the Hamiltonian of the form
\be
\hat{{\bf H}}
\to
\hat{{\bf H}}' = \hat{{\bf H}}
-
\varepsilon
\log_2 (\Xi |\Psi|^2)
,
\ee
and
the formal (non-thermal) temperature which can be
associated with this kind of entropy
is given by
$
T_\Psi
\equiv
(k_B \beta)^{-1} = (\partial E' / \partial S_\Psi )_\Xi
=
\varepsilon / ( k_B \ln 2 )
,
$
where
$E' = \langle \Psi | \hat{{\bf H}}' | \Psi \rangle
= E + \zeta T_\Psi S_\Psi $
is the total energy of the system  \cite{Yasue:1978bx,Zloshchastiev:2009zw}.
Rewriting $\varepsilon $ in terms of $\beta$, we recover
LogSE in  our notations (\ref{e-xmain}).
%, up to a sign of $\beta$.
For stationary states one can write it in the form
\be
%\hat{{\bf H}}' | \Psi \rangle  =
\left[
\hat{{\bf H}}
-
\beta^{-1} \ln (\Xi |\Psi|^2)
\right]
 \Psi  = E'  \Psi
,
\ee
whereas  the free energy is given by $E =
\langle \Psi | \hat{{\bf H}} | \Psi \rangle = E' - \zeta T_\Psi S_\Psi $.
Unlike the free energy, the energy $T_\Psi S_\Psi$
is engaged in handling the information $I_\Psi$
and thus unavailable to do dynamical work.

The \schrod equations of this type are suitable for describing
subsystems in which the information is not conserved
but being exchanged with environment, and where
one can introduce some sort of temperature and entropy.
Therefore, they can not be naively applied to the systems
%with conserved information and
without any kind of irreversibility
hence the negative results of
the experiments \cite{logexp} are not surprising.
On the other hand,
some Bose liquids might be good candidates, as being studied in this paper.
Another candidate
%for the theory which would need the logarithmic term
would be the yet unknown theory of
quantum gravity
where the debates still continue
about the black hole thermodynamics
and a possible information loss  \cite{Preskill:1992tc}.
%The quantum-gravitational applications of LogSE are given in Refs. \cite{Zloshchastiev:2009zw,Zloshchastiev:2009aw}.
To conclude, we write down the most important properties of LogSE:
\bit
\item Separability of noninteracting subsystems (as in the linear theory):
the solution of the LogSE for the composite system is a product
of the solutions for the uncorrelated subsystems;
\item Energy is additive for noninteracting subsystems (as in the linear theory);
\item Planck relation holds as in the linear theory;
\item All symmetry properties of the many-body wave-functions with respect to
permutations of the coordinates of identical particles are preserved in time,
as in the linear theory;
\item Superposition principle is relaxed to the weak one: the sum of solutions
with negligible overlap is also a solution;
\item Free-particle solutions, called \textit{gaussons}, have the coherent-states form,
%Gaussian functions which become
and upon the Galilean boost they become the uniformly moving
Gaussian wave packets modulated by the de Broglie plane waves;
\item Expressions for the probability density and current are the same
as in the linear theory.
\eit
All these properties except the last one and, perhaps, second last and third
last ones, are unique to LogSE
among all other local nonlinear \schrod equations.
Besides, many of these features are pertinent to the linear
\schrod equation which makes the logarithmic one a ``minimal'' nonlinear
modification in a sense.

\def\AnP{Ann. Phys.}
\def\APP{Acta Phys. Polon.}
\def\CJP{Czech. J. Phys.}
\def\CMPh{Commun. Math. Phys.}
\def\CQG {Class. Quantum Grav.}
\def\EPL  {Europhys. Lett.}
\def\IJMP  {Int. J. Mod. Phys.}
\def\JMP{J. Math. Phys.}
\def\JPh{J. Phys.}
\def\FP{Fortschr. Phys.}
\def\GRG {Gen. Relativ. Gravit.}
\def\GC {Gravit. Cosmol.}
\def\LMPh {Lett. Math. Phys.}
\def\MPL  {Mod. Phys. Lett.}
\def\Nat {Nature}
\def\NCim {Nuovo Cimento}
\def\NPh  {Nucl. Phys.}
\def\PhE  {Phys.Essays}
\def\PhL  {Phys. Lett.}
\def\PhR  {Phys. Rev.}
\def\PhRL {Phys. Rev. Lett.}
\def\PhRp {Phys. Rept.}
\def\RMP  {Rev. Mod. Phys.}
\def\TMF {Teor. Mat. Fiz.}
\def\prp {report}
\def\Prp {Report}

\def\jn#1#2#3#4#5{{#1}{#2} {\bf #3}, {#4} {(#5)}} %PRD
%\def\jn#1#2#3#4#5{{#1}{#2} {#3} {(#5)} {#4}}   %PLB style
% #1 tittle  #2 ser  #3 vol  #4 page  #5 year

\def\boo#1#2#3#4#5{{\it #1} ({#2}, {#3}, {#4}){#5}}
%\def\boo#1#2#3#4#5{ #1 ({#2}, {#3}, {#4}){#5}}  %PLB style
% #1 tittle  #2 publisher  #3 place  #4 year  #5 page/, p.789/

%\def\jn#1#2#3#4#5{{#1}{#2} {\bf #3}, {#4} {(#5)}}
% #1 tittle  #2 ser  #3 vol  #4 page  #5 year
%\def\boo#1#2#3#4#5{{\it #1} ({#2}, {#3}, {#4}){#5}}
% #1 tittle  #2 publisher  #3 place  #4 year  #5 page/, p.789/

%\newpage

\newpage

%\begin{figure}[htbt]
%\begin{center}\epsfig{figure=chempot-notrap.eps,width=  1.05\columnwidth}\end{center}
%\caption{
%Chemical potential for the logarithmic BEC with
%positive $\beta$
%in absence of any trap
%as a function of the dimensionless parameter $a/a_\beta = a/\sqrt{A_\beta}$,
%for different values of
%$N /e^3 \pi^{3/2}$.
%The dotted line corresponds to the value
%at which the potential vanishes at $a = a_\beta$.
%}
%\label{f:chempot-notrap}
%\end{figure}

%\begin{figure}[htbt]
%\begin{center}\epsfig{figure=energy-notrap.eps,width=  1.05\columnwidth}\end{center}
%\caption{
%Energy per particle for the logarithmic BEC with
%positive $\beta $
%in absence of any trap
%as a function of the dimensionless parameter $a/a_\beta$,
%for different values of
%$N /e^4 \pi^{3/2}$.
%The dotted line corresponds to the value
%at which the energy vanishes at $a = a_\beta$.
%}
%\label{f:energy-notrap}
%\end{figure}

%\begin{figure}[htbt]
%\begin{center}\epsfig{figure=chempot-trap.eps,width=  1.05\columnwidth}\end{center}
%%\includegraphics{fig1.ps}
%\caption{
%Chemical potential for the logarithmic BEC in an isotropic harmonic trap
%as a function of the dimensionless coupling $\beta^{-1}/\hbar \omega$,
%for different values of
%$N (a/a_\omega)^3/(\pi e)^{3/2}$.
%The dotted line corresponds to the value below (above) which the potential has
%no local maximum for strictly negative  (positive) $\beta$.
%}
%\label{f:chempot-trap}
%\end{figure}

\end{document}